\documentclass[journal]{IEEEtran}
\usepackage{subfigure}
\usepackage{cite}
\usepackage{hyperref}
\usepackage{csquotes}
\usepackage{balance}
\usepackage{color}
\usepackage{subfigure}
\usepackage{graphicx}
\usepackage{amsmath}
\usepackage{lipsum}

\usepackage{subfigure}
\usepackage{cite}
\usepackage{hyperref}
\usepackage{csquotes}
\usepackage{balance}
\usepackage{color}
\usepackage{subfigure}
\usepackage{graphicx}
\usepackage{tabularx}
\usepackage{lipsum}

\begin{document}

\title{Towards Industrial Private AI: A two-tier framework for data and model security}
\author{Sunder Ali Khowaja$^\dagger$, \IEEEmembership{Member IEEE}, Kapal Dev*$^\dagger$\thanks{$^\dagger$Joint first authors, with equal contributions to this paper}, \IEEEmembership{Member IEEE}, Nawab Muhammad Faseeh Qureshi*, \IEEEmembership{Senior Member IEEE}, Parus Khuwaja, and Luca Foschini, \IEEEmembership{Senior Member IEEE} % <-this % stops a space
\thanks{$^\dagger$Joint first authors, with equal contributions to this paper}
\thanks{*Corresponding author}%:
\thanks{Sunder Ali Khowaja is with Faculty of Engineering and Technology, University of Sindh, Jamshoro, Pakistan, and Department of Mechatronics Engineering, Korea Polytechnic University, Republic of Korea. Email: sandar.ali@usindh.edu.pk, sunderali@kpu.ac.kr}
% <-this % stops a space

\thanks{Kapal Dev is associated with the Department of Computer science, Munster Technological University, Ireland and Institute of Intelligent Systems, University of Johannesburg, South Africa, e-mail: (kapal.dev@ieee.org).}% <-this % stops a space

\thanks{Nawab Muhammad Faseeh Qureshi is associated with the Department of Computer Education, Sungkyunkwan University, Seoul, Korea., e-mail: faseeh@skku.edu.}
\thanks{Parus Khuwaja is with University of Sindh, Jamshoro. (e-mail: Parus.khuwaja@usindh.edu.pk).}

\thanks{Luca Foschini are associated with University of Bologna, Italy. e-mail: (luca.foschini@unibo.it)}
}


% \markboth{Journal of \LaTeX\ Class Files,~Vol.~14, No.~8, August~2015}%
% {Shell \MakeLowercase{\textit{et al.}}: Bare Demo of IEEEtran.cls for IEEE Journals}

\maketitle

\begin{abstract}
 With the advances in 5G and IoT devices, the industries are vastly adopting artificial intelligence (AI) techniques for improving classification and prediction-based services. However, the use of AI also raises concerns regarding data privacy and security that can be misused or leaked. Private AI was recently coined to address the data security issue by combining AI with encryption techniques, but existing studies have shown that model inversion attacks can be used to reverse engineer the images from model parameters. In this regard, we propose a Federated Learning and Encryption-based Private (FLEP) AI framework that provides two-tier security for data and model parameters in an IIoT environment. We propose a three-layer encryption method for data security and provided a hypothetical method to secure the model parameters. Experimental results show that the proposed method achieves better encryption quality at the expense of slightly increased execution time. We also highlight several open issues and challenges regarding the FLEP AI framework’s realization.
\end{abstract} 

\section{Introduction}\label{sec:intro}
With various ubiquitous devices connected through base stations, access points, and Internet of Things (IoT) gateways, Industrial IoT (IIoT) has brought a technological shift to how modern industries acquire and share their data. The large-scale deployment of IoT nodes and devices provides the analysts with enormous amounts of data to develop intelligent decision-making systems %. Although the IoT devices are capable of sharing huge volumes of data, the devices themselves are constrained with the computational complexity and storage capacity, therefore, the devices need a liaison for storing, managing, and effectively analyzing the data. The emergence of edge computing technologies and industrial clouds enabled the IoT devices to not only store huge amounts of data but also perform analysis 
useful for enterprise managers, technicians, operators, assembly line personnel, CEOs, and government personnel \cite{Usman2020}.
%Recently, industrial clouds leverage the use of artificial intelligence (AI) to deliver generalized and personalized services to the users on their smartphones such as restaurant recommendations, route directions, weather prediction, and so forth. 
However, this AI revolution in the high-tech industry comes at the cost of private data sharing and a proposition that the user is willing to share it with industrial clouds. At the user level, this issue is signified due to the sensitivity and privacy of the shared data such as minute-to-minute location, health records, and sequenced genome. Similarly, when applied to industries, the data sharing might consist of activities being performed inside the factory, new product details, trade secrets, formulas, and warehouse images. Thus, the data sharing raises security concerns as the sensitive information could be disclosed or leaked while sending it to the cloud \cite{Ster2022,Khowaja2021}. Examples of cyberattacks during 2021 or the leakage of location data using IBM weather services is a testimony of data security concerns%. A study conducted by Identity Theft Research Center in 2018 reported the breach of around 4.5 million sensitive records 
\footnote{https://www.idtheftcenter.org/wp-content/uploads/2019/02/ITRC\_2018-End-of-Year-Aftermath\_FINAL\_V2\_combinedWEB.pdf}. % that alone raises serious concerns over data security and privacy.The consequences of such information disclosure vary from loss of employment to death in the worst-case scenario. Furthermore, the industries are responsible for the security and privacy of users’ data that is collected, stored, and analyzed while working or being in the workplace premises. 
Industries also need to share the data with 3rd party firms for the analytical services. In such cases, the chances of data being compromised rise by multiple magnitudes.\\
The quid pro quo of data analytics associated with industrial clouds leads to a number of challenges with data security and privacy being the top one. %It has been suggested by many studies that the data needs to undergo an end-to-end encryption process before it is uploaded to the cloud. 
Furthermore, with the extended connectivity to the industrial clouds, the systems are not confined to a single user and that leads to one-to-many authorization issues \cite{Usman2019}. Even though one overcomes the data encryption issues, the attacks can also target the trained models that store the data or training parameters by manipulating, inverting, or stealing them. Moreover, complete solution in terms of data security and model privacy in the context of IIoT and industrial cloud is still missing, though some partial solutions are appearing.\\
Recently, the concept of Private AI was introduced \cite{Lauter2021} that employs encryption techniques to protect data privacy, however, the concept was abstract and only focused on one-tier privacy preservation, i.e., while sending it to 3rd party analysts. %We extend the concept of private AI for providing two-tier data security and privacy that not only secures the shared data but also preserves the trained AI models as shown in Figure 1. 
In this regard, we propose federated learning and encryption-based Private (FLEP) AI framework for data security and model privacy in an IIoT environment. To the best of our knowledge, this is the first study to propose two-tier data and model security framework using Private AI for IIoT environment. We also propose a three-layer data security method to encrypt images that can be decrypted at the 3rd party analyst’s end. The framework is general, therefore, any encryption algorithm that fits to the desired application can be employed at the data security level. We also propose the use of Federated learning (FL) and Homomorphic encryption (HE) to ensure the security of model parameters when transferred to the AI cloud. The contributions of this work are as follows:
\begin{itemize}
\item A FLEP AI framework is proposed for data security and model privacy in an IIoT environment.
\item A three-layer encryption method is proposed for data security.
\item Implementation flow for using HE with FL is provided.
\item Open challenges and issues concerning FLEP AI in IIoT environment are presented.
    \end{itemize}
The rest of the paper is structured as follows: Section 2 consolidates the literature review concerning the proposed work. Section 3 provides the details for each of the phases in the proposed FLEP AI framework. Section 4 presents experimental results to show the efficacy of our proposed work. Section 5 highlights open issues and challenges and Section 6 concludes the work, accordingly.
\begin{figure*}[h]
\centering
  \includegraphics[width=\linewidth]{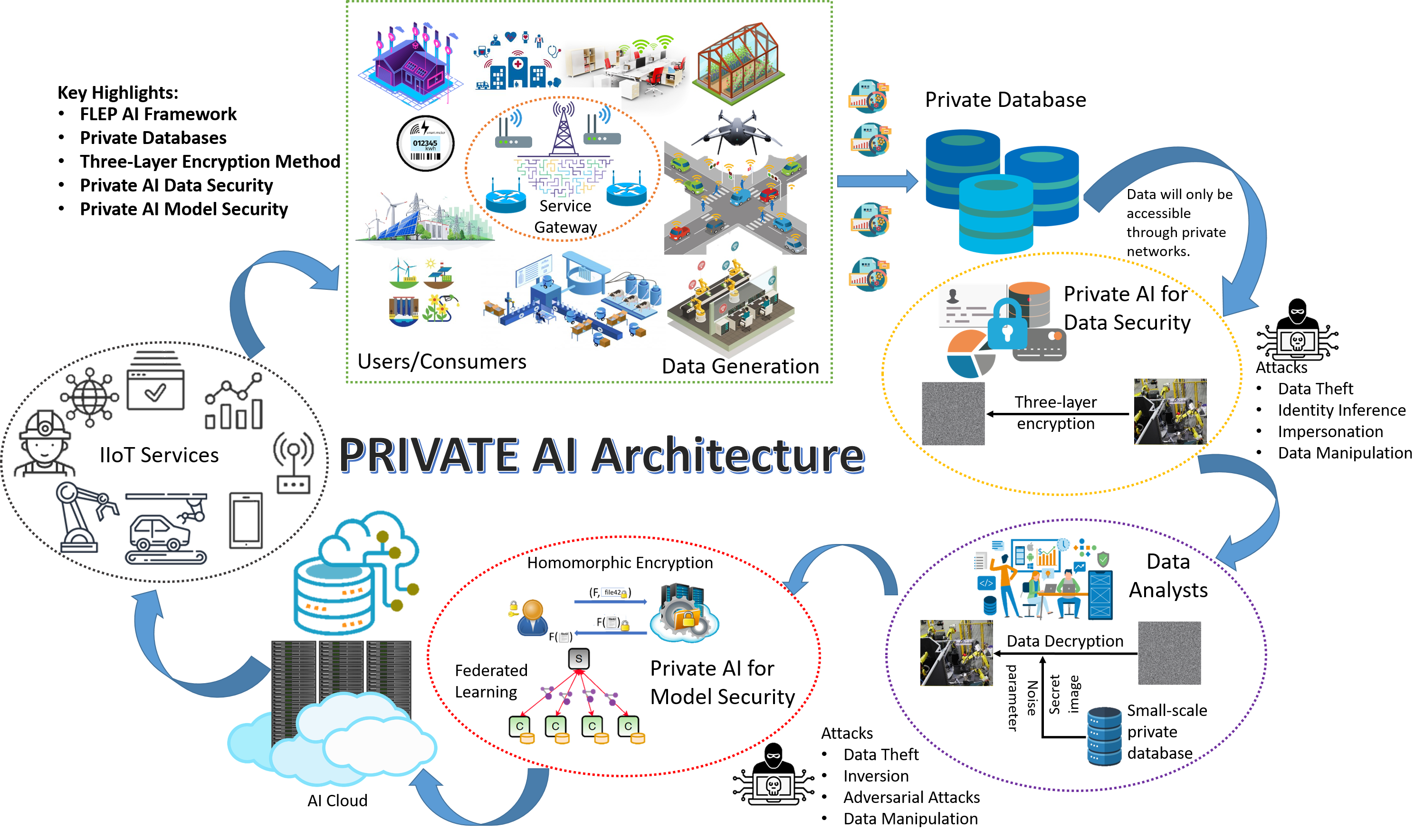}
  \caption{FLEP AI framework for two-tier data security in an IIoT environment.}
  \label{Fig1}
\end{figure*}

\section{Related Works}
Private AI refers to the coupling of encryption-based techniques applied to the enterprise data while training, classifying, or predicting from the samples. The private AI retains the privacy and security of the organizational data. The term was coined in 2019 referring to the use of HE to protect the data acquired from IoT devices. Recently, a Private AI collaborative research center has been established by Intel, Avast, and Borsetta in consortium with 8 top universities in the world which further supports the growing interest in this field\footnote{https://www.schunter.org/blog/2021/03/11/launch-of-the-private-ai-center/}.\\
The motivation of Private AI was derived from the work reported in \cite{Bachrach2016} that demonstrated the feasibility of using predictions derived by neural networks on HE data. A whitepaper on crypto \cite{Lauter2021} in International Association for Cryptologic Research explains the underlying technologies used for CryptoNets \cite{Bachrach2016} and explores the real-world applications associated with Private AI. Kaissis et al. \cite{Kaissis2020} did not explicitly used the term Private AI, but proposed data security and privacy preservation method using FL for medical images. Knott et al. \cite{Knott2021} proposed a secure multi-party computation method that could derive inferences from encrypted data, i.e. CRYPTEN. The focus of the study was predicting phonemes from audio modality while keeping the data private. The studies \cite{Boemer2020, Meng2020} proposed a privacy-preserving machine learning methods that use Ngraph and HE techniques to cope with the leakage of feature maps. The aforementioned works still focus on the data leakage, but when passed across intermediate convolutional layers. Subsequently the studies \cite{Usman2019, Kaissis2020} proposed the privacy preserving methods for medical-related data, \cite{Xian2020, Ma2017, Wang2019, Wang2021} image data, \cite{Usman2020, Ster2022, Khowaja2021} industry-related data, and \cite{Boemer2020, Meng2020, Zhu2020} inferential data, respectively. To the best of our knowledge, this is the first work that takes into account the security and privacy for both the data and the model while proposing a systematic Private AI framework. 

\section{FLEP AI framework}
The proposed FLEP AI framework is illustrated in Figure 1. The framework comprises seven components: data generation, private database, private AI for data security, data analysts, private AI for model security, AI cloud, and IIoT services. Some of the components are similar to a traditional IIoT network such as the data generation, data analysts, AI cloud, and IIoT Services. The uniqueness of the FLEP AI framework lies in the private database, private AI for data and model security components, accordingly. As for the main workflow, the information is acquired first from data generation which is then stored at private database suggesting that only private networks or the trusted parties with secret key information can access it. The data undergoes an encryption process at Private AI for data security. We propose a three-layer encryption method to show its effectiveness. The encrypted data is shared with data analysts that decrypt it with the help of opted noise level and secret image. The trained model and analysis is shared with the subsequent module to enforce privacy on model parameters. The approaches such as FL, HE, or combination of both can be applied to Private AI for model security stage. The trained model(s) from one or more data analysts are sent to AI cloud for further aggregation. The analytical results are then sent to service layer for activating IIoT explicit services or updating the aggregated model for the users as an implicit service. This section covers each of the phases and discusses them in detail.
\subsection{Data generation}
The data generation component in the FLEP AI framework comprises IoT devices used in multiple domains. For instance, smartphones and wearable sensors %such as photoplethysmography (PPG), electrodermal activity (EDA), electromyogram (EMG), electroencephalogram (EEG), and other sensors 
are the IoT devices mostly used in the healthcare domain. However, the same set of sensors can be used in an industrial environment to monitor the health of a worker ubiquitously as these sensors can be embedded in wearable devices. Inertial measurement units (IMUs) are mostly associated with movement recording, and can be used in a variety of domains including smart homes, smart healthcare, smart office, smart cities, smart industries, factories, and so forth. Meanwhile, LIDAR, pressure, infrared, and other similar sensors can be used for surveillance, water level monitoring, smart metering, greenhouses, and smart agriculture domains. The data acquired from all the aforementioned sensors (time-series, text, video, or speech) flow through wired or wireless means that are part of the data generation module. 
\subsection{Database}
In this study, we emphasize the use of private databases rather than the ones used publicly or on the cloud. The data acquired from the generation module is stored in a private database. A private database refers to a network setup where the data is acquired from the IoT devices and stored through virtual networks, thus, no public subnets are used. The private database ensures that the data can only be accessed from within the industrial tenancy via virtual networks. The database access from public endpoints is completely ceased with this type of network configuration. The advantages of using the private database are as follows:

\begin{itemize}
    \item It only needs a service gateway to connect the database with IoT devices.
    \item No need of transit routing setup.
    \item Capacity of attaining data security requirements.
\end{itemize}
However, the setup for creating a private database requires following resources:
\begin{itemize}
\item Database should reside within the virtual network region.
\item The virtual network should be configured with a private subnet.
\item There should be at least one network security group within the virtual network. 
\end{itemize}
A virtual firewall is created by the network security group that enforces safety rules to ensure the database security. The number of network security groups can be extended depending on the available resources for increasing the security level. 

\subsection{Private AI for data security}
%This work extends the scope of private data by proposing a framework that can not only preserve the data but also the trained models’ privacy and security. 
The FLEP AI framework allows any encryption method to be used on the data before sending it to the data analysts, but to show the module’s realization we propose a chaotic sequence and wavelet-transform-based data encryption method as shown in Figure 2. In this study, we discuss the implementation and design of the proposed method with respect to the image data. The private AI for the data security module can be divided into three segments: encryption, encoding, and adding noise. The keys used in generation of chaotic sequence along with the secret image and noise parameters will be used to decode and decrypt the information at the data analysts’ module.\\
We adopt the sub-block chaotic algorithm proposed in \cite{Xian2020} for scrambling the image. The method uses the matrix width and length to calculate the sub-blocks and apply the spiral scan transformation \cite{Xian2020} to scramble the pixel in sub-blocks. A chaotic sequence is then calculated and each sub-block is re-organized in ascending order. While applying the chaotic sequences, group keys are generated along with the sub-blocks' side length. These parameters will be stored at the private database and shared with data analysts’ module to descramble the image, accordingly. Once the image is scrambled, we apply discrete wavelet transform (DWT) to encode the image. The use of DWT not only provides an edge for data security but also compresses the data that can be beneficial for efficient use of bandwidth. We used the scrambled and a secret image, and applied 2D-discrete wavelet transform at level 1 to both the images and performing alpha blending, accordingly. The usage of a secret image ensures that even the image is descrambled, the information will not be completely leaked or restored. The third segment is the noise injection in the image that adds an extra layer of security to the data. The image is corrupted using additive white Gaussian noise using mean and standard deviation. The selected values for injecting random noise in the image will be used to select the deep learning model trained specifically for the selected noise levels. The purpose of using the encryption technique is to keep the data safe from pseudonymization attacks that can replace the original data with the artificially generated one and data theft.

\begin{figure*}[h]
\centering
  \includegraphics[width=\linewidth]{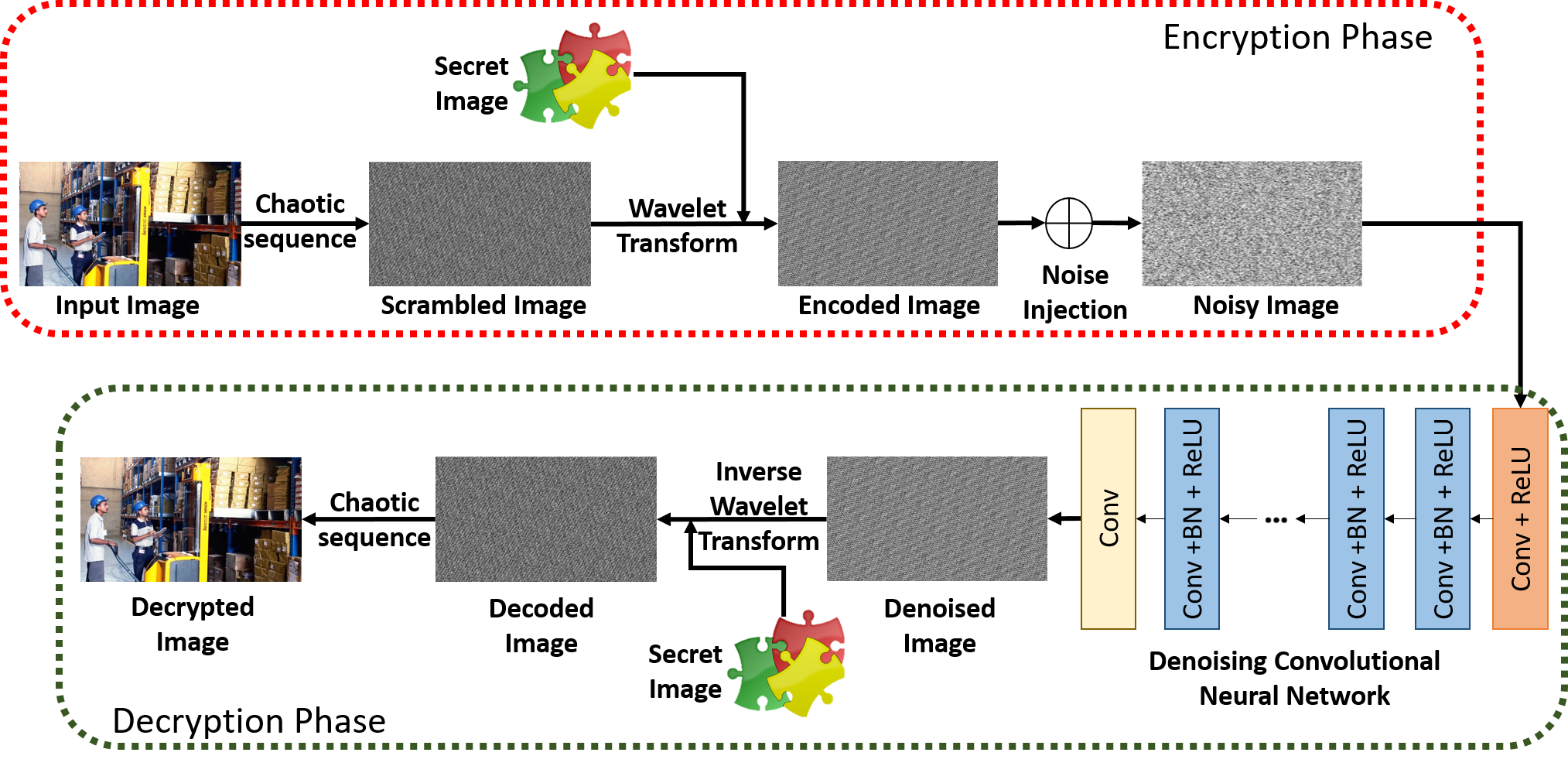}
  \caption{The proposed encryption and decryption pipeline for data security at Private AI and Data Analysts' phase.}
  \label{Fig1}
\end{figure*}

\subsection{Data Analysts} 
The data analysts phase in the FLEP AI framework has a small-scale private database that consists of private keys, secret images, and parameters such as mean and standard deviation of injected noise. Following the presumption, the data analysts will receive noisy image which is processed through a Denoising Convolutional Neural Network (DCNN). %The depth of DCNN is set to 9. The first layer consists of convolutional and rectified linear unit (ReLU) activation and the last layer consists of the only convolutional unit. The remaining 7 layers in the middle are comprised of convolutional, batch normalization (BN), and ReLU activation units. Each of the aforementioned layers extracts 64 feature maps. To ensure consistency in input and feature map size, zero paddings are employed. The network used an ADAM optimizer with default values, the size of the mini-batch was set to 64, and all the other hyperparameters including learning rate were optimized through the grid search method. The network was trained on the Waterloo exploration database having 4.744K images \cite{Ma2017}. We only trained the network with noise images having zero mean and the value of 25 as the standard deviation. The network was trained on MATLAB R2020a on a Core™i5 PC equipped with GeForce GTX 1080Ti GPU. 
The use of DCNN ensures that the data received is safe from noise attacks. Once the denoised image is obtained, the decryption phase performs inverse discrete wavelet transform (IDWT) with the same secret image used in the encryption process that gives us the decoded image. The decrypted image is then obtained by descrambling the image using the group key sequences (chaotic), sub-block side length, and spiral transformation. The proposed process adds three layers of security, however, the process can be made more or less secure depending on the data characteristics, sensitivity, and application requirements. The data analysts can perform the required analysis, extract patterns, and complete other required tasks on the decrypted data.   

%\begin{figure*}[h]
%\centering
%  \includegraphics[width=1.0\textwidth, height=275]{3.jpg}
%  \caption{Federated learning scheme without and with encryption. The scheme with encryption needs public or private keys to unlock the model parameters.}
%  \label{Fig1}
%\end{figure*}

\subsection{Private AI for Model Security}
One of the motivations to propose the FLEP AI framework is to extend the security footprint to model parameters as well. %The security of model parameters is of equal importance as its manipulation can change the classification or prediction outcome, thus, might lead to data leakage or information theft. 
Traditional machine learning approaches store the parameters and the data that not only requires larger storage capacity but also makes it vulnerable to data-related attacks. FL, a fully decentralized algorithm acquires copies of locally trained models from various users without storing the data. The algorithm performs further iterations to improve the model’s performance (global model), and sends back the parameters to the users \cite{Kaissis2020, Ster2022}. The benefit of using a FL approach is that data remain with the users, therefore, FL solves the issues regarding data ownership and governance to some extent. %Due to the privacy-preserving characteristics, the next generation communication and IIoT systems widely prefer to use a federated learning approach. 
Despite the aforementioned benefits, existing works suggest that a standalone technique does not guarantee model parameter security. The learning approach is susceptible to tampering, leakage, and unsecure aggregation which is unacceptable for many industrial applications \cite{Meng2020, Zhu2020}. Although, the FL approach does not store data, studies have shown that the data might be reconstructed with model inversion attacks \cite{Kaissis2020, Zhu2020}. Thus, FL needs to be combined with encryption techniques that could help fortifying the model parameters.

\begin{figure*}[h]
\centering
  \includegraphics[width=\linewidth]{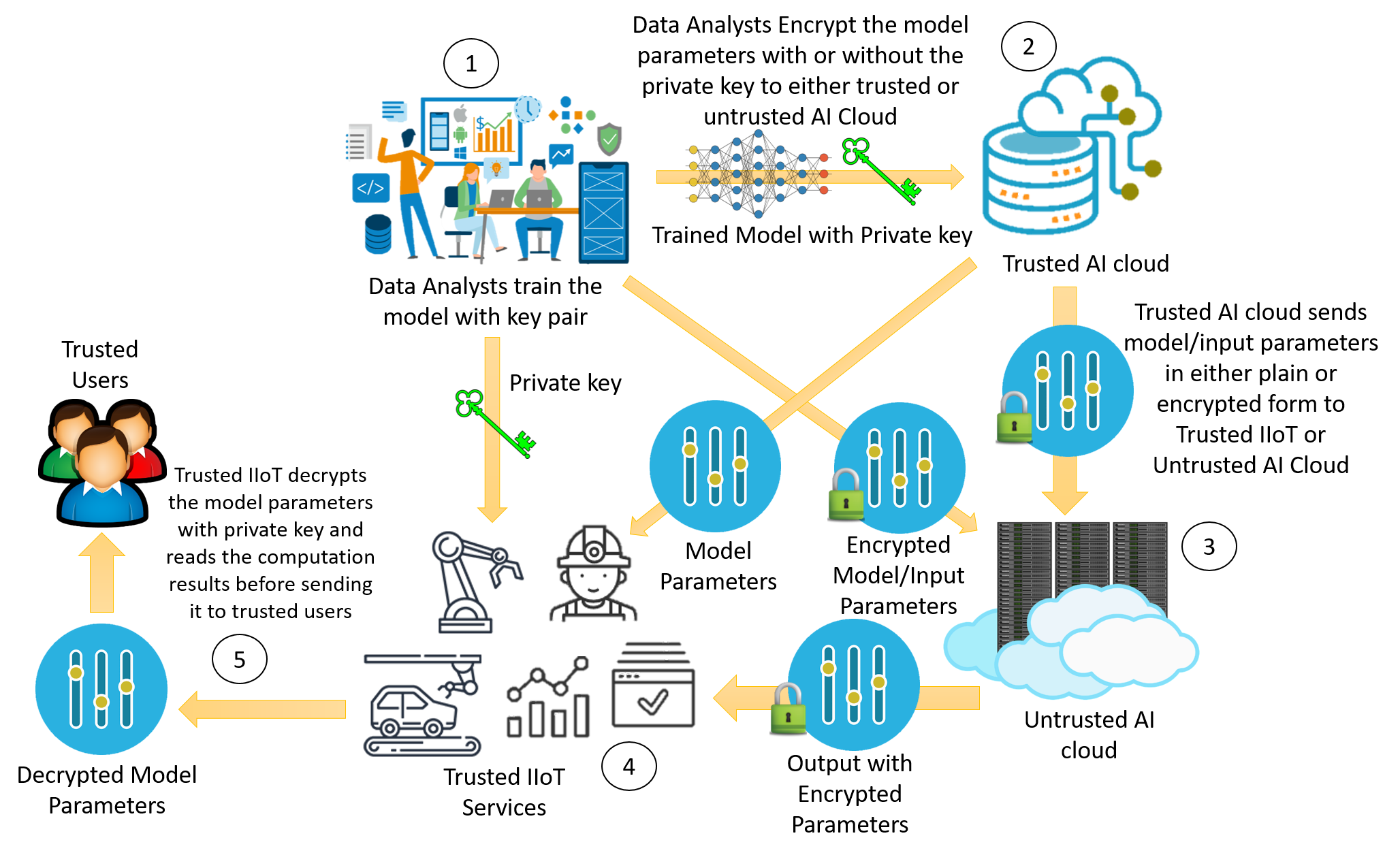}
  \caption{Homomorphic encryption process for federated learning approach.}
  \label{Fig1}
\end{figure*}
%The encryption algorithms have been widely used for the security of data and model parameters. Furthermore, the encryption algorithms are trusted by both the users, experts, and practitioners. 
For the sake of generality, the FLEP AI framework allows any encryption algorithm to be used in that phase. However, performing computation on the encrypted data has been a challenging task and yet remains an open research issue. In this regard, we suggest the use of HE technique that considers the encrypted data as plain text and allows its computation, processing, and analysis, accordingly \cite{Lauter2021}. %The said encryption technology is based on a mathematical concept, i.e., homomorphism, that preserves the structure while processing the data. %Homomorphic encryption only allows the mathematical operations that are homomorphic such as addition and multiplication, therefore, advanced encryption schemes neither can be considered here nor can be applied to the domain of neural networks [9]. Over the years, several studies have proposed homomorphic encryption algorithms that can be applied to data as well as model parameters.
This study proposes the use of HE to secure the model parameters and updates generated at the data analyst phase, and received by the AI cloud phase. An example of a HE process for FL approach is shown in Figure 3. Although, there might be multiple data analysts that build a shared model using FL while updating it in a periodic manner, but for explaining the process in Figure 3, we use a case of single data analyst with a locally trained model. Private AI assumes that the updates for both global and local model are only released exclusively to participating organizations or sections, therefore, the paired keys with the model will only be available to the said trusted parties, accordingly. The data analyst generates a paired public/private key with encrypted model parameters. The keys are shared with the trusted AI cloud and IIoT services for accessing the model parameters. However, if needed, the HE module sends the trained model parameters as a plaintext or encrypted form to untrusted AI cloud for further computation without any paired key. Lets suppose the plaintext is denoted by $t_1$ and $t_2$. The HE allows the untrusted AI cloud to perform operations on the encrypted text denoted by $E(t_1)$ and $E(t_2)$. The result of the operation will yield a new ciphertext that would be sent to the the trusted IIoT services, with an assumption that the operation or content can be recovered by summing the two plaintext, i.e. $t_1 + t_2$. The IIoT services, then, can access the parameters using the corresponding key. The trusted users then receive the output from implicit IIoT services, accordingly.

\subsection{AI cloud and IIoT Services}
AI cloud works at larger scale in terms of data analysis, data storage, model aggregation, and model update. %Let’s assume a company like Amazon having privateAI setups for multiple markets or industries. Each industry has its own privateAI for data security. Some of the industries might share the same data analysts while other approach to different organizations. All of the data analysts perform the analysis and train a local model coupled with homomorphic encryption with key pairs. The AI cloud collects all the models from different data analysts trained locally, aggregates them to generate a global one, and shares the data back to the analysts for updating their local models.
The AI cloud can reside in parent companies, headquarters, or with governments. The access to the model parameters at this phase also depends on the trust between data analysts and AI cloud owners as the trusted partner can be given a public key while the untrusted partner has to perform the model aggregation on encrypted parameters, accordingly. Based on the decision forwarded by the AI cloud, the services will be activated or an action would be performed inline, accordingly. The IIoT services include, but are not limited to manufacturing, monitoring tools, automation, safety, quality inspections, industrial apps, and actuators. These services are either directly provided or sent to the users in the form of analysis.

%\section{Use Case Scenario}
%\textcolor{blue}{Let’s assume a company like Amazon having private AI setups for multiple markets or industries for this hypothetical scenario. Each industry has its own private database and private AI for data security. Some of the industries might share the same data analysts while other approach to different organizations. The use of Private AI data security helps to cope with data theft, identity inference, impersonation, and data manipulation. Therefore, if any data analyst is impersonated or tried to steal the data, they need to have the secret image or private key to decrypt it. All of the data analysts perform the analysis and train a local model coupled with HE and key pairs. The AI cloud collects all the models from different data analysts trained locally, aggregates them to generate a global one, and shares the data back to the users for updating their local models through the IIoT service module. In case the model is sent to an untrusted AI cloud, it would be anyway safe from data theft, model inversion attacks, and data manipulation due to the encryption technique being applied. It is also essential to cope with the security concerns at IIoT service module as some of the services require the use of untrusted 3rd party applications. The proposed Private AI architecture provides multiple layers of security in order to preserve the privacy of data as well as model.} 

\section{Experimental Results for data encryption}
In order to prove the efficacy of the proposed algorithm (private AI for data security), we perform the encryption on three different images as suggested in \cite{Wang2019}. The reason for choosing these three images is two-fold. The first is that many studies have used these images as a standard to perform encryption and the second is to use these images for a fair comparison. The original files are single-channel grayscale images having the size 256x256. We use the sub-block size of 2x2, the depth of DCNN is set to 9. The first layer consists of convolutional and Rectified Linear Unit (ReLU) activation and the last layer consists of the only convolutional unit. The remaining 7 layers in the middle are comprised of convolutional, Batch Normalization (BN), and ReLU activation units. Each of the aforementioned layers extracts 64 feature maps. To ensure consistency in input and feature map size, zero paddings are employed. The network used an ADAM optimizer with default values, the size of the mini-batch was set to 64, and all the other hyperparameters including learning rate were optimized through the grid search method. The network was trained on the Waterloo exploration database having 4.744K images \cite{Ma2017}. We only trained the network with noisy images having zero mean and the value of 25 as the standard deviation. The network was trained on MATLAB R2020a on a Core™i5 PC equipped with GeForce GTX 1080Ti GPU. The original, encrypted, and decrypted images are shown in Figure 4. \\
In this study, we evaluate the quality of encryption using histograms, Number of Pixels Change Rate (NPCR) and Unified Average Changing Intensity (UACI), information entropy, encryption quality, and total time for encryption and decryption, respectively. These measures are commonly used to evaluate the encryption method when differential attacks are applied. The histograms are also shown in Figure 4 along with their corresponding images. It is suggested that the histogram should be balanced and its visualization should be closer to uniform distribution for improved encryption quality \cite{Wang2019,Wang2021}. The results show that the proposed method generates a close uniform distribution at its encryption phase. We summarize the results obtained using the proposed method and its comparison with the study \cite{Wang2019} in Table 1. The values obtained for NPCR are very close to the values proposed in \cite{Xian2020} that suggests that the change in the pixel values has occurred due to the encryption process. The obtained results for UACI are also close to the optimum values. The information entropy measure provides information regarding the complexity and randomness of encrypted images. The optimum value of information entropy is suggested to “8” in existing works \cite{Wang2019}, therefore, a value closer to 8 should be considered better in terms of encryption. The encryption quality is computed from the difference of intensity values between the original and the encrypted image. A higher value for this measure is considered to be better in terms of encryption quality. Lastly, the encryption times are considered to be of vital importance. The numbers on the lower side are preferred suggested that the numbers for all the aforementioned measures are also satisfactory. The results show that the proposed method not only achieves optimum results but also surpasses the performance recorded in \cite{Wang2019}. The proposed method yields slightly higher execution time due to the added security (Three-layer security method). However, the number of security layers can be reduced depending on the employed application and the industry requirements to improve the execution time. For instance, the removal of noise addition and denoising block reduces the encryption time to 0.98 seconds for three images on average but slightly affects the encryption quality. Therefore, we can assume that there is a trade-off between the computational complexity and the encryption quality of the system. Furthermore, it should also be noted that the addition of noise provides an efficient coping mechanism for pseudonymization and noise attacks \cite{Xian2020,Wang2019}. In addition, we carry out a comparative analysis on benchmark dataset\footnote{http://sipi.usc.edu/database/} and various standard images used in \cite{Xian2020,Wang2021} with respect to NPCR and UACI values to show the effectiveness of the proposed approach. There are variations in terms of image size in the employed dataset, therefore, we increase the sub-block size by a power of 2 as the image size increases. All other parameters are the same for this experiment. The aforementioned datasets provdies an opportunity to perform a fair comparison with the existing works to showcase the efficacy of the proposed work. Table 2 reports the obtained results. It can be noticed that the proposed method not only achieves better results from both the methods (near to average values in the range \cite{Wang2021}) but also exhibits the least standard deviation suggesting that the encryption quality does not deviate much while varying image size and intensities, accordingly.

\begin{figure*}[h]
\centering
  \includegraphics[width=\linewidth]{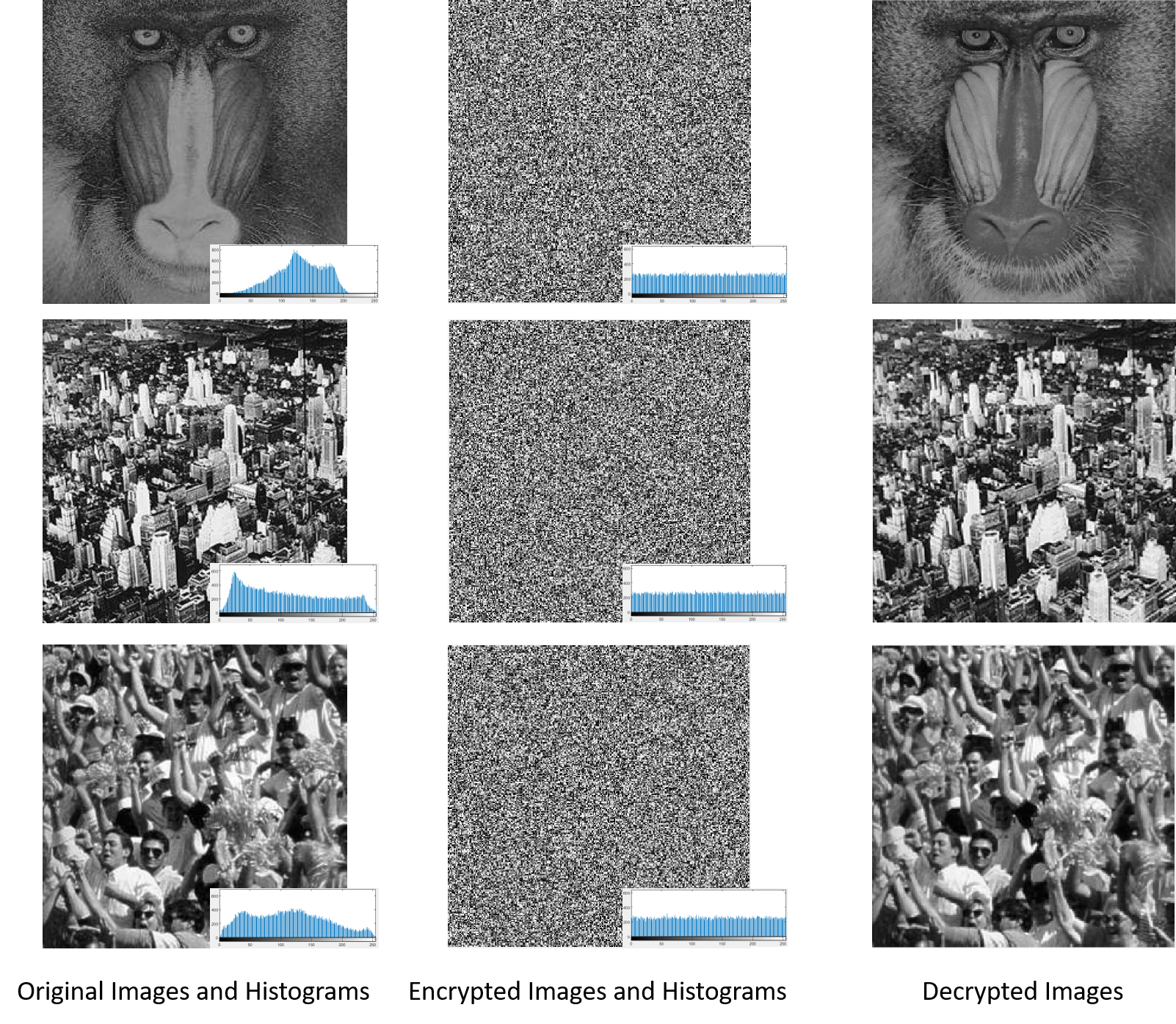}
  \caption{Encrypted and decrypted images using the proposed three-layer data security method.}
  \label{Fig1}
\end{figure*}

\begin{table}[]
\caption{Comparative analysis of image encryption quality on three standard images with \cite{Wang2019}}
\begin{tabular}{|l|l|l|l|}
\hline
Evaluation Metric     & Baboon Image & City Image & Fan Image  \\ \hline
\multicolumn{4}{|c|}{{[}13{]}}                                 \\ \hline
UACI                  & 32.0312      & 35.2089    & 31.7543    \\ \hline
NPCR                  & 99.5845      & 99.5232    & 99.6167    \\ \hline
Encryption Quality    & 61.4676      & 45.3828    & 35.8046    \\ \hline
Information Entropy   & 7.9572       & 7.9514     & 7.9563     \\ \hline
Total Encryption Time & 1.0485 sec   & 1.0520 sec & 1.0540 sec \\ \hline
\multicolumn{4}{|c|}{Ours}                                     \\ \hline
UACI                  & 33.4319      & 33.5087    & 33.4815    \\ \hline
NPCR                  & 99.5857      & 99.5896    & 99.6295    \\ \hline
Encryption Quality    & 72.197       & 57.458     & 42.396     \\ \hline
Information Entropy   & 7.9973       & 7.9971     & 7.9971     \\ \hline
Total Encryption Time & 1.5228 sec   & 1.4936 sec & 1.5302 sec \\ \hline
\end{tabular}
\end{table}

% \usepackage{multirow}
\begin{table}[]
\caption{Comparison with state-of-the-art on Miscellaneous image dataset using UACI and NPCR.}
\label{tab:my-table}
\begin{tabular}{|l|l|l|l|l|l|l|}
\hline
\multicolumn{1}{|c|}{{File Name}} & \multicolumn{3}{c|}{UACI (33.282,33.645)} & \multicolumn{3}{c|}{NPCR (99.57,99.60)} \\ \cline{2-7} 
\multicolumn{1}{|c|}{}                           & {[}11{]}      & {[}14{]}      & Ours      & {[}11{]}      & {[}14{]}     & Ours     \\ \hline
5.1.09     & 33.482 & 33.456 & 33.464 & 99.61  & 99.61  & 99.58  \\ \hline
5.1.10     & 33.478 & 33.451 & 33.456 & 99.62  & 99.61  & 99.58  \\ \hline
5.1.11     & 33.428 & 33.483 & 33.457 & 99.63  & 99.61  & 99.59  \\ \hline
5.1.12     & 33.409 & 33.453 & 33.462 & 99.60  & 99.60  & 99.58  \\ \hline
5.1.13     & 33.446 & 33.436 & 33.462 & 99.59  & 99.61  & 99.58  \\ \hline
5.1.14     & 33.455 & 33.485 & 33.483 & 99.62  & 99.61  & 99.59  \\ \hline
5.2.08     & 33.488 & 33.509 & 33.458 & 99.62  & 99.61  & 99.59  \\ \hline
5.2.09     & 33.490 & 33.508 & 33.465 & 99.61  & 99.61  & 99.58  \\ \hline
5.2.10     & 33.483 & 33.450 & 33.465 & 99.61  & 99.60  & 99.58  \\ \hline
5.3.01     & 33.490 & 33.455 & 33.459 & 99.60  & 99.61  & 99.58  \\ \hline
5.3.02     & 33.454 & 33.446 & 33.458 & 99.61  & 99.61  & 99.58  \\ \hline
7.1.01     & 33.432 & 33.478 & 33.462 & 99.62  & 99.60  & 99.59  \\ \hline
7.1.02     & 33.438 & 33.417 & 33.457 & 99.62  & 99.61  & 99.58  \\ \hline
7.1.03     & 33.416 & 33.466 & 33.459 & 99.60  & 99.61  & 99.58  \\ \hline
7.1.04     & 33.509 & 33.446 & 33.461 & 99.61  & 99.61  & 99.58  \\ \hline
7.1.05     & 33.488 & 33.457 & 33.459 & 99.60  & 99.61  & 99.58  \\ \hline
7.1.06     & 33.418 & 33.488 & 33.460 & 99.60  & 99.62  & 99.58  \\ \hline
7.1.07     & 33.455 & 33.444 & 33.459 & 99.61  & 99.61  & 99.58  \\ \hline
7.1.08     & 33.395 & 33.449 & 33.454 & 99.61  & 99.62  & 99.59  \\ \hline
7.1.09     & 33.499 & 33.481 & 33.457 & 99.60  & 99.61  & 99.58  \\ \hline
7.1.10     & 33.512 & 33.485 & 33.460 & 99.60  & 99.61  & 99.58  \\ \hline
7.2.01     & 33.483 & 33.456 & 33.459 & 99.61  & 99.61  & 99.58  \\ \hline
boat.512   & 33.403 & 33.489 & 33.454 & 99.61  & 99.61  & 99.58  \\ \hline
gray21.512 & 33.494 & 33.493 & 33.461 & 99.62  & 99.61  & 99.58  \\ \hline
ruler.512  & 33.517 & 33.455 & 33.464 & 99.62  & 99.61  & 99.59  \\ \hline
Mean       & 33.463 & 33.466 & 33.461 & 99.61  & 99.61  & 99.58  \\ \hline
STD        & 0.0368 & 0.0231 & 0.0056 & 0.0096 & 0.0045 & 0.0044 \\ \hline
\end{tabular}
\end{table}

\section{Open Issues and Challenges}
The proposed FLEP AI framework is specifically designed to increase data and model parameter security. Accordingly, this section will solely focus on the issues and challenges associated with the data and model parameter security.\\
\textit{\textbf{Operating and Computation cost:}} Private AI is relatively a new field, therefore, employing the architectural components proposed, such as private database would not be a practical choice for industries having limited data. Recently, private 5G networks are on the rise which might help in decreasing the operating cost of private AI through secure network slices.\\
It has been proven by the existing studies that the performance of classification/prediction tasks is heavily affected by the encrypted data \cite{Lauter2021,Kaissis2020}. In this regard, the design of encryption algorithms that does not affect the performance would remain the point of concern. Furthermore, adding encryption also adds a necessary computational overhead on the system to protect the data from theft and leakage that eventually increases the computation cost.\\ %Recently Amazon Web Services (AWS) proposed the privacy preserving machine learning (PPML) \cite{Meng2020} that is able to perform computation on the encrypted data which provides a way forward to deal with the computational cost while employing privacy preserving techniques. \\ %The results will be forwarded back to the user for decryption. Their results revealed that the computational cost can be reduced to some extend by employing the PPML technique.} \\
%\textit{\textbf{Noise modeling:}} The selected domain for the FLEP AI framework is an IIoT environment that might involve devices as small as a wearable sensor to as large as a bank of machinery. Moreover, there are a lot of machines in factory environments and manufacturing plants that generate noise which can affect the data sent to the private database. This work takes into account AWGN noise, however, other types of noises such as Rayleigh, Laplacian, Rician, and others can also be generated in an IIoT environment and can corrupt the data samples before they are stored in a private database. In this regard, a pre-processing phase can cope with the noise modeling issue for multiple data modalities.\\
\textit{\textbf{Heterogeneous data modalities:}} This study focuses on the image data for providing a kind of realization to private AI-based systems. In IIoT environments, the data types might vary from texts, speech, time-series data to images and videos. Each of the data types needs different techniques for encryption to be employed in order to yield good performance. The heterogeneity of data modalities can also affect the private databases as a uni-modal system can have devices with different sampling rates, standardized units, and performance issues. The data from heterogeneous devices will need pre-processing techniques before sending it to the data storage.\\
\textit{\textbf{Hard-delay constraints:}} It is evident from the experimental results that encryption takes time. For the systems demanding fast response or having hard real-time characteristics, the encryption techniques need to be improved such that the total encryption and decryption time is reduced. However, such techniques might make the private AI system vulnerable to certain attacks. \\
\textit{\textbf{Scalable AI:}} Scalable AI refers to the capability of AI infrastructure, data, and algorithm to operate at certain complexity, speed, and size admissible for a large-scale real-time system. In order to develop a scalable system, the challenges of data scarcity, labeling of training samples, reusability, and parallel processing needs to be undertaken. %Recently, a study \cite{So2020} proposed a scalable approach that allows data-owners to jointly train AI models while keeping the data private. 
In FLEP AI framework, the idea can be extended to multiple data analysts companies training AI models in parallel while keeping the data private. Furthermore, the data encryption tasks can reuse the secret images, keys, wavelet coefficients, and noise parameters to enhance the reusability factor, thus, increasing the scalability of the system.  \\
\textit{\textbf{Evolution of attacks:}} Over the years, the attacks on data as well as model parameters have been evolved and became more efficient. The algorithms and systems also need to be evolved at the same pace in terms of encryption techniques, use of virtual networks, and public/private keys. Recently, a study \cite{Zhu2020} proposed the reconstruction of data from intermediate feature maps or gradients of a trained network. Such attacks needs to be taken into consideration, especially in the case of IIoT where data falls under the government legislation or commercially valuable. In the case of the FLEP AI framework, the evolution should be performed in terms of scrambling technique, usage of more than one secret image, and using different values of mean and standard deviation for noise injection. \\
%\textit{\textbf{Private user devices:}} This work focuses on securing the data and model parameters once they are acquired from the devices in the deployed network. Existing studies suggest that most of the attacks are performed at the user device level. If the data is corrupted at the acquisition level, then the analysis would eventually result in losses. An authorization mechanism needs to be devised or a private user protocol should be proposed so that the devices having proper consensus can be used for data acquisition.

\section{Conclusion}
Data analytics in an IIoT environment has witnessed great advances in terms of AI methods. However, the issues of data privacy and security remain at large. The unregulated use of data can be misused, thus, might lead to monetary or health losses. Furthermore, it has been suggested that data can be reconstructed from the model parameters through model inversion attacks. This paper proposes a private AI-based framework, i.e. FLEP AI, that provides two-tier security for data and model parameters, accordingly. We proposed a data encryption technique that adds three layers of security to the data before undergoing data analysis. We also provided a realistic framework for integrating FL approach with HE to secure the model parameters before sending them to the AI cloud. Our experimental results for data encryption showed that the method can achieve good encryption quality while slightly increasing the computation time. The study also highlights open challenges and issues that can be faced while implementing the FLEP AI framework. \\
As a future specific case, lets assume a company like Amazon having Private AI setups for multiple industries. We assume that each industry is equipped with Private database(s). The use of Private AI data security module ensures that the data is safe from theft, manipulation, impersonation, and identity inference. Some of the industries might share the same data analysts while other use 3rd party services. These analysts train the model locally on the data acquired and decrypted from Private AI data security module. The trusted AI cloud collects models from data analysts, aggregates them, and generate a global one, which is then shared to the users through IIoT service module. In some cases, some IIoT services require 3rd party services. In this regard, the Private AI model security uses HE to keep the model safe from the likes of parameters leakage and model inversion attacks. Based on the decrypted parameters and decisions, a specific service is initiated or the local model is updated for the end-users, accordingly.\\   
This study only implements Private AI data security module in Flep AI framework. We are keen to extend the work by employing multi-modal data and a model-security method to exhibit the proof-of-work concerning the proposed framework. This will allow us to not only address the highlighted open issues but also find some new research directions in the context of private AI.\\

%is an abstract depiction and needs to be explored further to design detailed architectural components. We are very keen to collaborate with some industries in the future for designing a proof-of-work concerning the FLEP AI framework.  

\section*{Acknowledgment}

The authors would like to thank all the contributors who shared data with us for this analysis. 
\bibliography{ref.bib}
\bibliographystyle{IEEEtran}
\vskip -0.1\baselineskip plus -1fil
\vspace{-0.25cm}
\begin{IEEEbiographynophoto} {Dr. Sunder Ali Khowaja} received the Ph.D. degree in Industrial and Information Systems Engineering from Hankuk University of Foreign Studies, South Korea.  He has served as an Assistant Professor at Department of Telecommunication Engineering, University of Sindh, Pakistan. He is currently associated with Department of Mechatronics Engineering, Korea Polytechnic University, Republic of Korea, in the capacity of postdoctoral research fellow. He is also serving as a reviewer for many reputed journals, including, IEEE Transactions on Industrial Informatics, IEEE Access, IEEE Internet of Things Journal, IEEE Transactions on Network Science and Engineering, IEEE Transactions on Medical Imaging, and others. He also served as a Technical Program Committee member in CCNC 2021, Mobicom 2021, and Globecom 2021 workshops. He is currently assisting in the capacity of Guest Editor at Computers and Electrical Engineering, Human-Centric Computing and Information Sciences, and Sustainable Energy Assessment and Technologies Journals. His research interests include Data Analytics, Deep Learning, and Communication Systems based applications.
\end{IEEEbiographynophoto}
\vskip -2\baselineskip plus -1fil
\vspace{-0.25cm}
\begin{IEEEbiographynophoto}{Kapal Dev} is Assistant Lecturer at MTU, Ireland and senior research associate at University of Johannesburg, South Africa. He is AE in Springer WINE, Elsevier PHYCOM, IET Quantum Communication, IET Networks, Topic Editor in MDPI Network. He is contributing as GE in Q1 journals; IEEE TII, TNSE, TGCN, Elsevier COMCOM and COMNET. He served(ing) as Lead chair in ICDCS 2022, MobiCom 2021, Globecom2021, $\&$ CCNC 2021 workshops. He contributed as PI for Erasmus + ICM, CBHE, and H2020 Co-Fund projects. His research interests include Blockchain, Wireless Networks and Artificial Intelligence. 
\end{IEEEbiographynophoto}
\vskip -2\baselineskip plus -1fil
\vspace{-0.25cm}
\begin{IEEEbiographynophoto}{Nawab Muhammad Faseeh Qureshi} Nawab Muhammad Faseeh Qureshi received Ph.D. degree in computer engineering from Sungkyunkwan University, South Korea. He is currently an Assistant Professor with Sungkyunkwan University, South Korea, where, he is actively involved in the Big Data Project and discusses new trends in IoT-enabled data analytics. His research interests include big data analytics, context-aware data processing of the Internet of Things, and cloud computing. 
He received the SAMSUNG scholarship for his Ph.D. degree and the Superior Research Award from the College of Information and Communication Engineering on account of his research contributions and performance during Ph.D. studies.
\end{IEEEbiographynophoto}
\vskip -2\baselineskip plus -1fil
\vspace{-0.25cm}
\begin{IEEEbiographynophoto}{Parus Khuwaja} is pursuing her Ph.D. degree in financial analytics from University of Sindh, Jamshoro. She is currently working as an Assistant Professor at University of Sindh, Jamshoro. Her interests include Data analytics, Machine learning for Ambient Intelligence, Stock Portfolios, and Financial securities.
\end{IEEEbiographynophoto}
\vskip -2\baselineskip plus -1fil
\begin{IEEEbiographynophoto}{Luca Foschini} received the graduation degree from the University of Bologna, Italy, and the Ph.D. degree in computer science engineering from the University of Bologna, in 2007, where he is an Associate Professor of computer engineering. He is currently working on mobile crowdsensing and crowdsourcing and management of cloud systems for smart city environments. His interests span from integrated management of distributed systems and services to wireless pervasive computing and scalable context data distribution infrastructures and context-aware services.
\end{IEEEbiographynophoto}
%\vskip -2\baselineskip plus -1fil
%\vspace{-0.25cm}
%\begin{IEEEbiographynophoto}{Paolo Bellavista} received MSc and PhD degrees in computer science engineering from the University of Bologna, Italy, where he is now a full professor of distributed and mobile systems. His research activities span from pervasive wireless computing to online big data processing under quality constraints, from edge cloud computing to middleware for Industry 4.0 applications. He serves on several Editorial Boards, including IEEE COMST (Associate EiC), ACM CSUR, and Elsevier JNCA and PMC. He is the scientific coordinator of the H2020 BigData project IoTwins.
\end{document}